\documentclass[prl,twocolumn,superscriptaddress,letterpaper]{revtex4}
\usepackage{graphicx,amssymb,bm,natbib,amsfonts,amsmath}
\usepackage{hyperref}
\usepackage{subfigure}
\begin{document}
\title{Interfacial Ferromagnetism in LaNiO$_3$/CaMnO$_3$ Superlattices}

\author{A. J. Grutter}
\affiliation{Department of Materials Science and Engineering, University of California, Berkeley, CA 94720, USA}
\affiliation{Materials Sciences Division, Lawrence Berkeley National Laboratory, Berkeley, CA 94720, USA}
\affiliation{Geballe Laboratory for Advanced Materials and Department of Applied Physics, Stanford University, Stanford, CA 94305, USA}
\author{H. Yang}
\affiliation{Department of Materials Science and Chemical Engineering, University of California-Davis, One Shields Avenue, Davis CA 95616, USA}
\author{B. J. Kirby}
\affiliation{NIST Center for Neutron Research, National Institute of Standards and Technology, Gaithersburg, MD, 20899}
\author{M. R. Fitzsimmons}
\affiliation{Los Alamos Neutron Science Center, Los Alamos National Laboratory, Los Alamos, NM 87545 USA}
\author{J. A. Aguiar}
\affiliation{Materials Science and Technology Division, Los Alamos National Laboratory, Los Alamos, NM 87544}
\author{N. D. Browning}
\affiliation{Chemical and Materials Sciences Division, Pacific Northwest National Laboratory, 902 Battelle Boulevard, Richland, WA  99352 USA}
\author{C. A. Jenkins}
\affiliation{Advanced Light Source, Lawrence Berkeley National Lab, Berkeley, CA 94720, USA}
\author{E. Arenholz}
\affiliation{Advanced Light Source, Lawrence Berkeley National Lab, Berkeley, CA 94720, USA}
\author{V. V. Mehta}
\affiliation{Department of Materials Science and Engineering, University of California, Berkeley, CA 94720, USA}
\affiliation{Materials Sciences Division, Lawrence Berkeley National Laboratory, Berkeley, CA 94720, USA}
\author{U. S. Alaan}
\affiliation{Department of Materials Science and Engineering, University of California, Berkeley, CA 94720, USA}
\affiliation{Geballe Laboratory for Advanced Materials and Department of Applied Physics, Stanford University, Stanford, CA 94305, USA}
\author{Y. Suzuki}
\affiliation{Department of Materials Science and Engineering, University of California, Berkeley, CA 94720, USA}
\affiliation{Materials Sciences Division, Lawrence Berkeley National Laboratory, Berkeley, CA 94720, USA}
\affiliation{Geballe Laboratory for Advanced Materials and Department of Applied Physics, Stanford University, Stanford, CA 94305, USA}

\date{\today}

\begin{abstract}
We observe interfacial ferromagnetism in superlattices of the paramagnetic metal LaNiO$_3$ and the antiferromagnetic insulator CaMnO$_3$. LaNiO$_3$ exhibits a thickness dependent metal-insulator transition and we find the emergence of ferromagnetism to be coincident with the conducting state of LaNiO$_3$. That is, only superlattices in which the LaNiO$_3$ layers are metallic exhibit ferromagnetism. Using several magnetic probes, we have determined that the ferromagnetism arises in a single unit cell of CaMnO$_3$ at the interface. Together these results suggest that ferromagnetism can be attributed to a double exchange interaction among Mn ions mediated by the adjacent itinerant metal.
\end{abstract}

\maketitle
Emergent phenomena at perovskite oxide interfaces have been studied intensively in the last decade in order to understand how mismatches in bands, valences, and interaction lengths give rise to novel interfacial ground states. Surprisingly, there have been only a handful of successful efforts demonstrating new magnetic ground states at interfaces. Among them is ferromagnetism (FM) attributed to the conductive layer at the interface between LaAlO$_3$ and SrTiO$_3$ that is associated with fractional charge transfer.\cite{Hwang}  In CaRuO$_3$/CaMnO$_3$ (CRO/CMO) superlattices, FM coupling has been attributed to interfacial double exchange.\cite{Takahashi} However, the nature of these new FM states is not yet well understood and the difficulty of isolating intrinsic interfacial effects from alloying or bulk phenomena remains an obstacle to our understanding of this interfacial FM.\cite{Mike}

In CRO/CMO superlattices, the interface FM is attributed to itinerant electrons in the CRO mediating a canted antiferromagnetic state among the Mn ions in CMO at the interfaces.\cite{Takahashi, Nanda, Chunyong,Freeland} However, interdiffusion may give rise to FM, since the solid solution CaRu$_x$Mn$_{1-x}$O$_3$ is FM for 0.1 $<$ x $<$ 0.7.\cite{Maignan} More recently, FM has been reported to be induced in typically paramagnetic LaNiO$_3$ by adjacent FM LaMnO$_3$ in LaMnO$_3$/LaNiO$_3$ superlattices.\cite{Marta} Although mechanistically very different, in both cases ultrathin strongly correlated metallic layers are essential for the generation of FM in these superlattices. In both systems, the strongly correlated metal is paramagnetic in bulk and appears to be on the verge of antiferromagnetism. In the case of LaNiO$_3$, it is generally agreed that the material is on the verge of a metal insulator transition (MIT) that can be induced by reducing its thickness.\cite{Scherwitzl,Boris,May} The origin of this thickness dependent MIT is not well understood but several mechanisms have been proposed.\cite{Marta,Scherwitzl,Munoz} Despite these open questions, epitaxial LaNiO$_3$ layers are excellent candidates for exploring the origin of interfacial ferromagnetism in systems where itinerant electrons may mediate ferromagnetic exchange. In particular, the metal insulator transition offers a unique tool for separating the effects of itinerant electrons at the interface from others such as intermixing, epitaxial strain, and defects.

In this paper, we report FM in LaNiO$_3$/CaMnO$_3$ (LNO/CMO) superlattices originating in the Mn ions and confined to one unit cell at the interface as determined by X-ray magnetic circular dichroism (XMCD) and polarized neutron reflectometry (PNR). The FM is highly dependent on the LNO metallicity. Metallic samples (LNO layer thickness $\geq$ 4 unit cells) show hysteresis in magnetization vs. applied field measurements while insulating samples (LNO layer thickness $<$ 4 unit cells) do not. Because FM occurs only in the presence of conducting LNO layers, we argue that interfacial double exchange, not intermixing or defects, is responsible for the FM ordering. We speculate that this double exchange originates in slight leakage of Ni e$_g$ electrons into the CMO layer. Such leakage is predicted to be on the order of 0.07 electrons per Mn in order to stabilize FM.\cite{Nanda} As the LNO decreases in thickness we suggest that the Ni e$_g$ electrons are localized, reducing the leakage and destroying the FM state.

We have grown and characterized high quality (n,m) superlattices where \emph{n} and \emph{m} are the number of LNO and CMO unit cells per layer, respectively. Unless noted otherwise, the number of superlattice repetitions was 8. The LNO/CMO superlattices were grown by pulsed laser deposition with a KrF excimer laser at 700 $^{\circ}$C in 4 Pa of O$_2$ on (100) oriented LaAlO$_3$ substrates. Atomic force micrographs show smooth terraced films with typical RMS roughnesses of 0.13-0.14 nm, on the order of half a unit cell.

\begin{figure}
\includegraphics[width=8.6 cm]{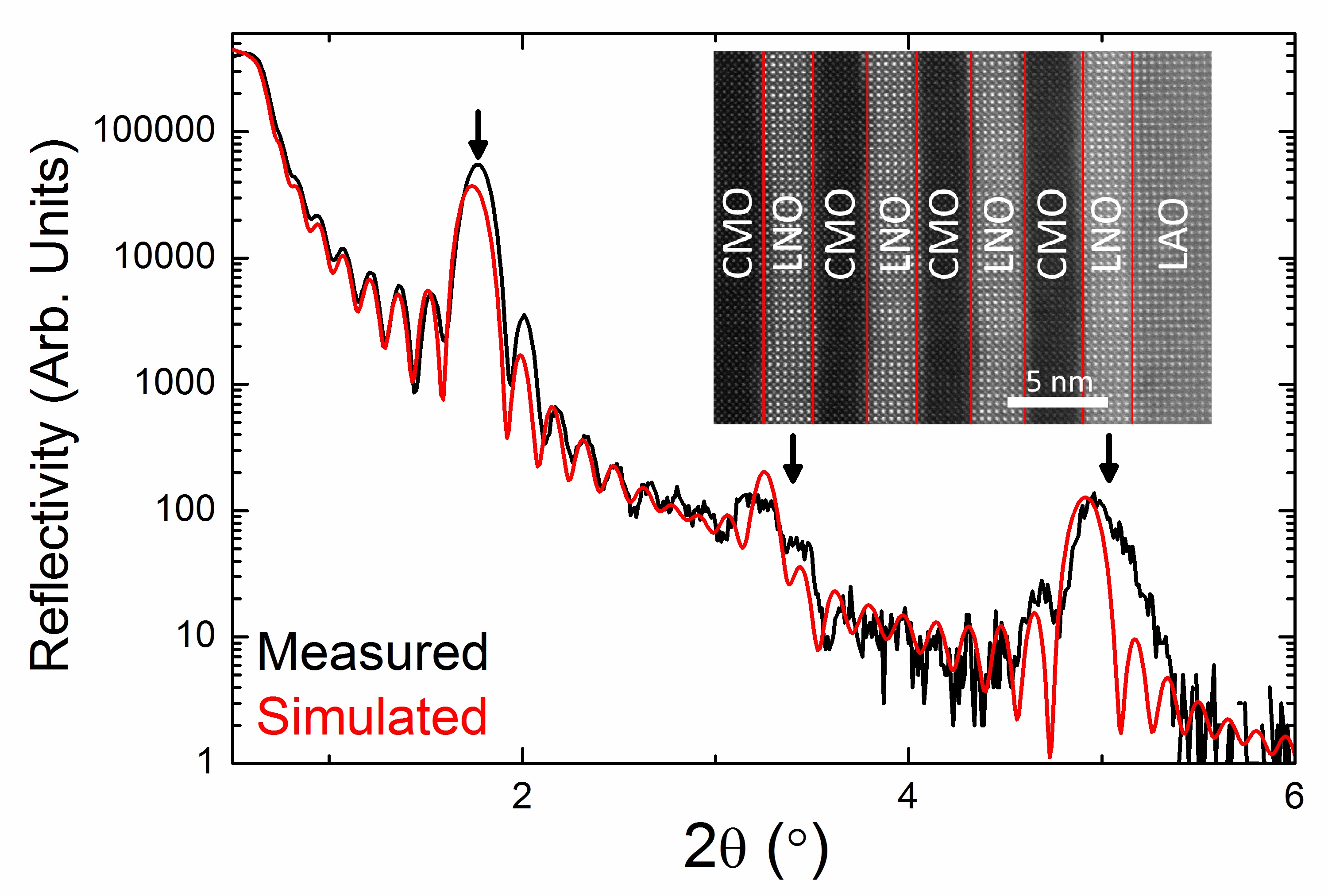}
\caption{(color online) X-ray reflectivity and theoretical fit of a (6,8)$_{10}$ film showing total thickness fringes as well as first, second, and third order superlattice reflections (marked by arrows). (Inset) High resolution STEM Z-constrast image of the same sample.}
    \label{fig:fig1}
\end{figure}

X-ray diffraction $\theta$-2$\theta$ scans (not shown) revealed high quality growth in the expected (100) orientation, while X-ray reflectivity (XRR) confirmed the existence of highly abrupt interfaces. A representative XRR of a (6,8)$_{10}$ sample is shown in Figure 1. The high frequency fringes correspond to the total thickness while the larger low frequency peaks correspond to the superlattice Bragg reflections. This spectrum is very well fit by a model with an interfacial roughness of 0.18 nm, suggesting very little mixing at the interface. Agreement between the expected total (53 nm) and measured (54.3 nm) superlattice thicknesses is within the expected error of the measurement, indicating that the deposited layer thicknesses closely match the intended value. The mosaic spreads of LaNiO$_3$ and CaMnO$_3$ films normalized to the substrate mosaic spread ($\Delta\omega_{film}/\Delta\omega_{LaAlO_3}$) were between 1.7-2.3, indicating high quality epitaxial growth.

To probe structural quality and interfacial abruptness more directly, we performed cross sectional high resolution scanning transmission electron microscopy (STEM) on a (6,8) superlattice. STEM shows that we have fabricated high quality films with excellent epitaxial registry across interfaces. Although small local variations in layer thickness are observed, the results are consistent with expected thicknesses determined through XRR measurements. Figure 1 (inset) shows a Z-contrast image illustrating the interfacial sharpness, with contrast abruptly switching across the interfaces and at most a single unit cell of intermixing. Although heavy La-doping of the CaMnO$_3$ is the primary mechanism by which interdiffusion might induce FM, the abrupt interfaces in Figure 1 (inset) suggest that interdiffusion is unlikely to result in an interfacial FM state.\cite{Schiffer}

To identify the magnetic ions, we obtained element specific magnetic information using X-ray absorption (XA) and XMCD measurements performed at beamline 6.3.1 of the Advanced Light Source. Measurements were performed in total electron yield mode at 20K. The field was parallel to the direction of X-ray propagation and both were at a 30$^{\circ}$ angle of incidence to the in-plane direction of the film. The incident light was maintained at a constant circular polarization. XA spectra of the Mn L$_{3,2}$ edge showed a 10.0 eV energy splitting between the L$_3$ and L$_2$ peaks and a L$_3$/L$_2$ peak ratio of approximately 2. These features suggest a Mn valence between 3.9-4.0+.\cite{Subias, Wang2, Schmid} The valence state is near the expected Mn$^{4+}$, but does not preclude a small amount of electron leakage at the interface, predicted by Nanda et al. to result in a valence of Mn$^{3.93+}$ in the interfacial layers of CRO/CMO superlattices.\cite{Nanda} XMCD was obtained from the difference in XA signal in $\pm$ 1.7 T T applied along the direction to X-ray propagation. The difference signal was normalized to its sum to obtain the data shown in Figure 2. XMCD indicates that the magnetic response arises from the Mn ions of the CMO layer (Figure 2). We do not observe any XMCD signal at the Ni L$_{3,2}$ edges, thus indicating that, to within experimental resolution, there is no magnetic response associated with the Ni ions in LNO. These observations conclusively demonstrate a magnetic response occurring only in the CMO layers.

We performed high-resolution electron energy loss spectroscopy (EELS) on the Mn L$_{3,2}$ edge to probe for spatial variation in the Mn valence state that would suggest FM induced by intermixing. The Mn valence was determined using both constrained multiple linear least squares fitting and L$_{3,2}$ peak heights, yielding oxidation states of 4$^+$ $\pm$ 0.3$^+$ and 3.75$^+$ $\pm$ 0.2$^+$, respectively.\cite{Tan, Varela} Neither technique showed a statistically significant difference in valence between Mn at the interface and in the middle of the CaMnO$_3$ layer, suggesting very little modification of the Mn valence through La-doping of the CaMnO$_3$ at the interface. Although a slight Mn valence modification through leakage of Ni e$_g$ electrons cannot be detected to within the error of the EELS measurements, we can eliminate a FM La$_x$Ca$_{(1-x)}$MnO$_3$ phase (0.5 $<$ x $<$ 1 corresponding to a Mn valence $\leq$ 3.5+). Thus, it is very unlikely that La-doping significant enough to induce FM has occurred.\cite{Schiffer} 

\begin{figure}
\includegraphics[width=8.6 cm]{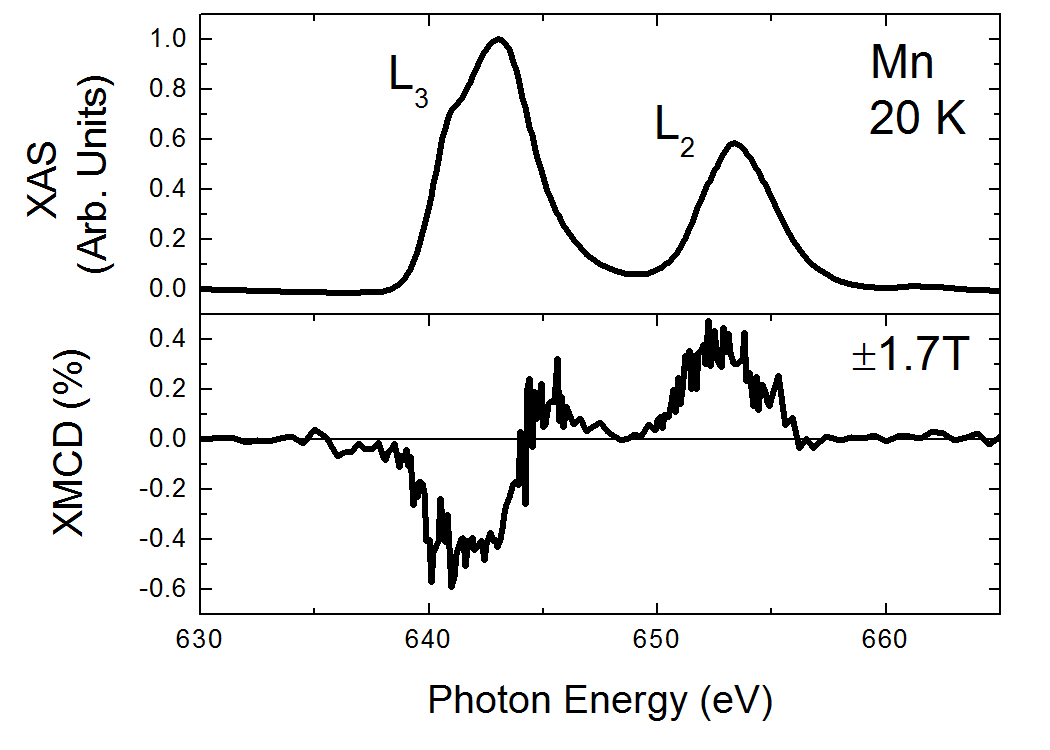}
\caption{a) X-ray absorption spectra of the Mn L$_{3,2}$ edge of a (6,8) superlattice. This spectra is consistent with a Mn valence between 3.9+ and 4.0+, very near bulk CMO. b) X-ray magnetic circular dichroism of the Mn L$_{3,2}$ edge showing asymmetry associated with a magnetic response in the Mn ions.}
\end{figure}

To probe the magnetic depth profile, we performed PNR on a (5,8) sample using the Asterix beamline at Los Alamos National Lab. The superlattice was cooled to 15 K in a 0.7 T magnetic field applied in the plane of the sample. Incident neutrons were polarized to be spin-up or spin-down with respect to this field. The specular reflectivity of spin-polarized neutrons is dependent on the depth profile of the nuclear composition and the depth profile of the sample magnetization component parallel to the applied field. Thus, sample magnetization manifests as a splitting of the spin-up and spin-down reflectivities. Such splitting is evident in Figure 3(a), which shows the spin-dependent reflectivities as functions of wavevector transfer along the surface normal (Q$_Z$) near the 1st order superlattice Bragg reflection. The measured reflectivity in Figure 3 is scaled by the theoretical reflectivity of the LaAlO$_3$ substrate. The PNR data were then fitted using the Refl1D software package.\cite{Brian} Figure 3(b) shows a model with periodic FM in the superlattice which is consistent with the data.

\begin{figure}
\includegraphics[width=8.6 cm]{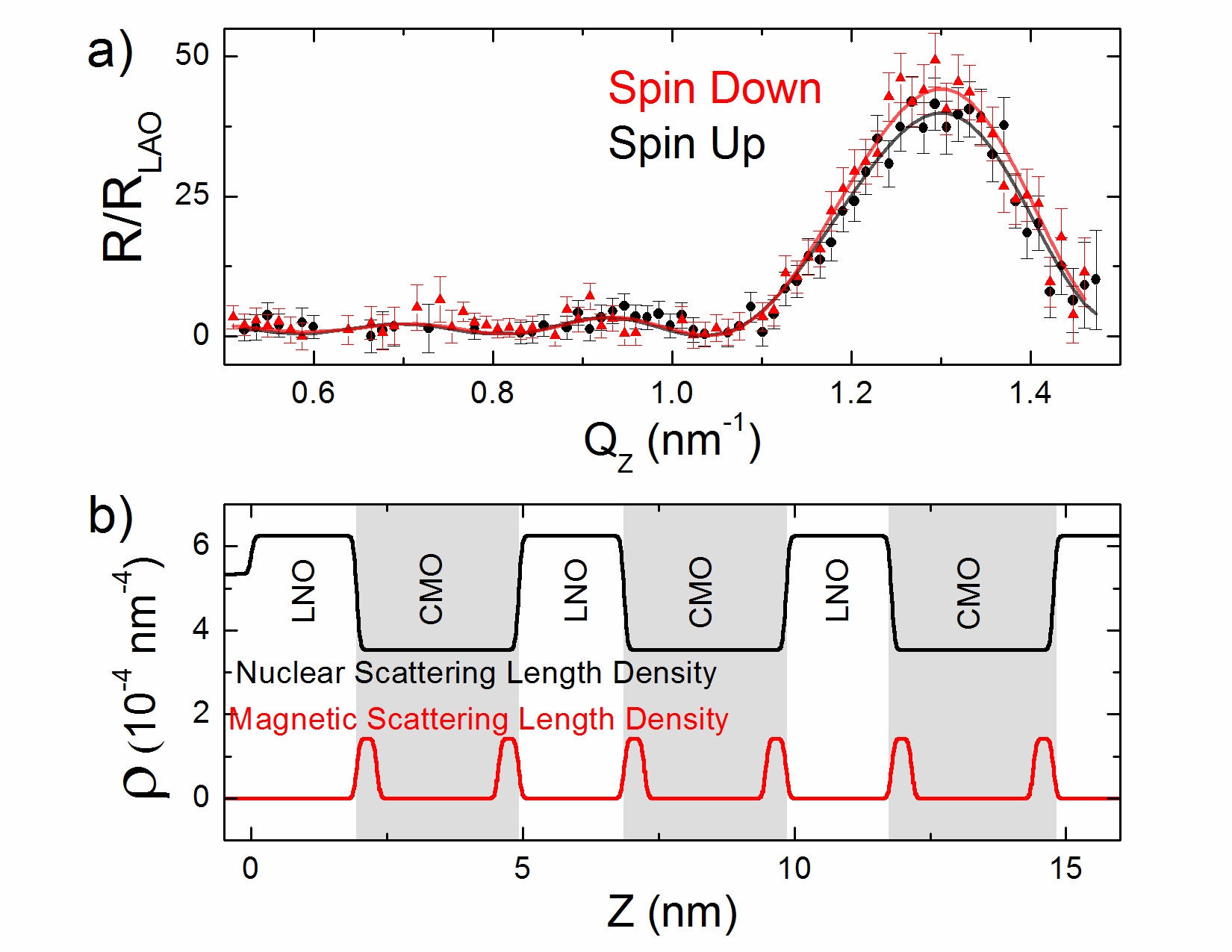}
\caption{(color online) a) Fitted polarized neutron reflectivity with standard error at 800 mT and 15 K at the first order superlattice reflection of a (5,8) superlattice. b) We show below the model used to obtain this fit, in which we assume one unit cell of magnetized CMO.}
\end{figure}

Constraints imposed by atomic force microscopy, XMCD, and XRR measurements require a model in which intermixing is limited to less than 2 \AA, all magnetism originates in the CMO layer, and layer thickness and nuclear scattering length density are within 10\% of expected values. Using this model, we find that only a magnetic depth profile in which the magnetization is confined to within one unit cell of the interface can reproduce the observed spectrum. As shown in Figure 3, the calculated reflectivity corresponding to this model accurately reproduces the spin-dependent Bragg reflection. All other possible thicknesses of the FM layer (2-4 unit cells) result in a reversal of the splitting on the first superlattice Bragg reflection. We conclude therefore that the FM in the CMO layer is confined to one unit cell at the interface.

SQUID magnetometry, performed at 10 K with $\pm$ 5 T fields applied parallel to the substrate surface (in-plane), shows hysteretic loops indicative of FM in superlattices where n $\geq$ 4 but not in those where n $<$ 4. Figure 4a shows magnetic moment vs. applied field for typical FM (2,8), (4,8), (6,8) samples. For magnetization vs. temperature measurements, samples were cooled to 10 K in a field of 1 T and magnetization measurements were taken while warming in a field of 0.01 T. The magnetization of FM films exhibits linear, non Curie-Weiss, temperature dependence with a clear T$_C$ of 75-80 K while non-FM n $<$ 4 films show no indication of a transition (Fig. 4(b) inset). No magnetic transitions were observed between 160-260 K, the range of T$_C$s expected for a ferromagnetic alloy of La$_x$Ca$_{1-x}$MnO$_3$.\cite{Schiffer} Assuming a model with a single magnetic monolayer of CMO at the interface, we find that the FM films saturated between approximately 0.5-1.0 $\mu_B$ per interfacial Mn (Fig. 4a). Saturated magnetic moment was independent of CMO thickness, as demonstrated in Figure 4b by a comparison of similar FM (6,8), (6,14), and (6,20) films, which all saturate at 0.5 $\mu_B$ per interfacial Mn. In addition, increased superlattice thickness for larger m results in a larger coercive field, which may be a result of a slight roughening of the superlattice with increasing overall film thickness.

As with all weak FM signals, contamination must be eliminated as a potential source. We note that only one temperature dependent magnetic transition is observed in magnetization vs. temperature scans. Additionally, no hysteresis is observed at temperatures above 80 K, well below the expected T$_C$s of  contaminants such as iron. Finally, we deposited an alloyed film of La$_{0.5}$Ca$_{0.5}$Ni$_{0.5}$Mn$_{0.5}$O$_3$ on LaAlO$_3$ and characterized it magnetically using SQUID magnetometry under the same conditions as the superlattice measurement. We found that the magnetism in the alloyed film is much too weak to explain the observed effects in superlattice films.

\begin{figure}
\includegraphics[width=8.6 cm]{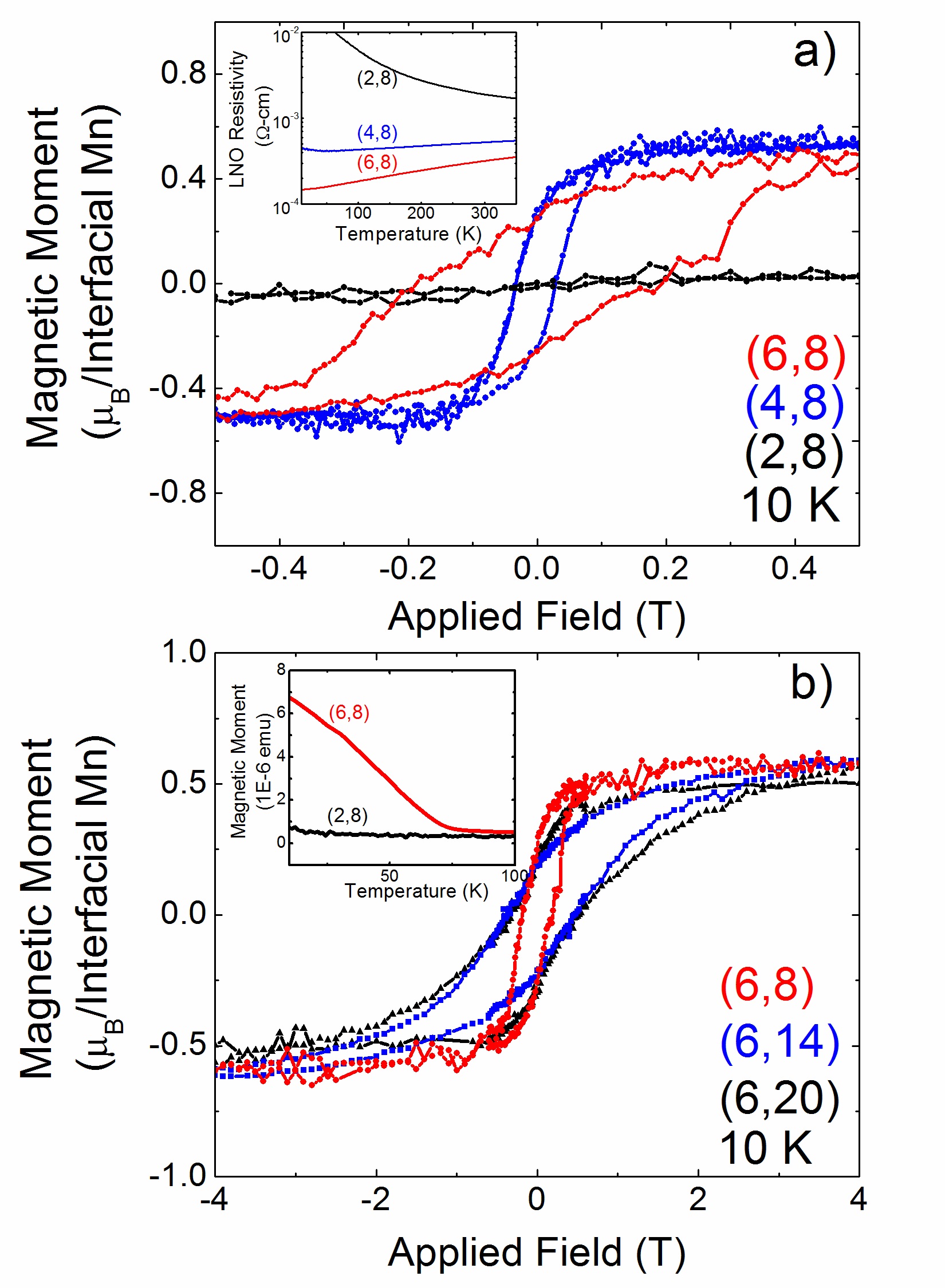}
\caption{(color online) a) Magnetic hysteresis loops for a series of typical (2,8), (4,8), and (6,8) superlattices. (Inset) Resistivity vs. temperature from 5-350 K for (2,8), (4,8), and (6,8) superlattices showing a thickness dependent metal-insulator transition. This transition coincides with the transition from FM to nonmagnetic  behavior of the films. b) Magnetic hysteresis loops showing scaling of the magnetic moment with the number of interfaces rather than CMO layer thickness. (Inset) Magnetic moment vs. temperature from 10-150 K. A FM transition is observed between 75-80 K only in n $\geq$ 4 superlattices.}
\end{figure}

Transport measurements taken in the van der Pauw geometry and varying temperature from 5-350 K show the expected thickness dependent metal insulator transition in the LNO layers. Samples for which n $\geq$ 4 were metallic. Assuming conduction only across the thickness of the LNO layers, we found the samples to have resistivities at 5K on the order of 1$\times10^{-4}$ $\Omega$-cm. These values are in good agreement with other examples of PLD grown LNO.\cite{May, Liu, Scherwitzl, Son} At LNO thicknesses of n = 2, the samples are insulating, showing an exponential temperature dependence indicative of thermally activated hopping conduction. Measured resistivities were comparable to those reported for other LNO films of similar thickness grown on (100) LaAlO$_3$. Figure 4(a) inset illustrates the transition, showing the conductivity of typical (2,8), (4,8), and (6,8) superlattices. The disappearance of FM as the thickness of the LNO layer is decreased is a strong indication that the FM is closely tied to the metallicity of the LNO layer. We theorize that mobile electrons in the Ni$^{3+}$ e$_g$ band extend into the interfacial CMO and mediate FM in the form of a double exchange interaction. As the LNO thickness is decreased, the Ni electrons no longer mediate FM ordering and the FM disappears.

Through this study, we have demonstrated interfacially confined FM in LNO/CMO superlattices and now address alternative explanations of the magnetism. Intermixing induced FM would be expected to persist through the LNO metal-insulator transition, while an interfacial effect in which mobile electrons from the LNO mediate FM in the CMO would be expected to be closely tied to the LNO conducting state. Intermixing induced FM would also be expected to increase with thicker CMO layers due to the greater deposition time resulting in increased intermixing. However, no such effect is observed. TEM and EELS show no evidence of intermixing at the levels required to induce a FM moment.\cite{Schiffer} Similarly, PNR measurements explicitly rule out uniform magnetization of CMO that might arise from oxygen vacancies or other defects. The T$_C$s of the materials are also inconsistent with both La-dopant and defect induced FM in CMO, which are expected to exhibit T$_C$s of at least 160-260K and 130K, respectively.\cite{Neumeier, Schiffer} The observed saturated magnetic moments are much too strong to be consistent with defect induced FM such as that observed in CMO nanoparticles.\cite{Markovich} In any case, magnetism arising throughout the CMO layer must scale with the CMO thickness, which we do not observe. 

Therefore we believe that the only remaining explanation is that of an interfacial magnetic interaction between the LNO and CMO which results in 1 unit cell thick FM layers as indicated by the PNR measurements. Such an exchange mechanism is likely analogous to that shown in CRO/CMO superlattices, in which it has been proposed that mobile electrons from CRO mediate canted FM in the CaMnO$_3$.\cite{Takahashi, Nanda, Chunyong} In this model, mobile Ni e$_g$ electrons leak into the first unit cell of the adjacent CMO layer, facilitating double exchange among the Mn ions. Unlike ferromagnetism corresponding to intermixing, such a small electron leakage is expected to result in only a very slight reduction of the Mn valence similar to that predicted for CRO/CMO.\cite{Nanda} This change is unlikely to be detected by either X-ray absorption or EELS measurements. A transition of the LNO to an insulating, potentially antiferromagnetic state in the superlattices with thin LNO layers results in localization of the electrons, a reduction in leakage, and the loss of the interfacial FM.

In conclusion, we have demonstrated FM in LNO/CMO superlattices that can only be explained in terms of an interfacial double exchange interaction. We find that LNO undergoes a metal-insulator transition as the LNO layer thickness is decreased. We observe FM with a T$_C$ of 70 K in the conducting superlattices but not in the insulating ones. We believe that a preponderance of evidence from SQUID magnetometry, XMCD, and PNR points to the FM originating in one unit cell of CMO at the interface. In particular, the strong dependence of the FM on the conducting state of LNO is indicative of an interfacial double exchange interaction mediated by the LNO e$_g$ band. 

This work was supported by the U.S. Department of Energy, Office of Basic Energy Sciences, Division of Materials Sciences and Engineering under Contract Nos. DE-AC05-76RL01830 (Berkeley \& ALS),  DE-SC0008505 (Stanford), and  DE-AC05-76RL01830 (H.Y.). Los Alamos National Laboratory is operated by Los Alamos National Security LLC under DOE Contract DE-AC52-06NA25396. U.S.A. is supported by the Office of Naval Research (N00014-10-1-0226). A portion of the PNNL research was performed using EMSL, a national scientific user facility sponsored by the Department of Energy's Office of Biological and Environmental Research.

\end{document}